\documentstyle[preprint,epsf,tighten,eqsecnum,amsfonts,aps,pre]{revtex}

\def\chichi{\left({\chi_m\over\chi_1}\right)}
           
	 \def\ci{{\cal I}}
\def\be{\begin{equation}}  \def\ee{\end{equation}}

\input epsf

\begin{document} \draft \preprint{}

\title{Undulation instability of lipid membranes under an electric field}

\author{Pierre Sens$^\dag$ \& H. Isambert$^\ddag$} \address{$\dag$Institut
Charles Sadron - 6 rue Boussingault, 67083 Strasbourg - France\\
$\ddag$L.D.F.C rue de l'Universit\'e, 67000 Strasbourg - France\\
Email:sens@ics.u-strasbg.fr}

\maketitle

\begin{abstract}
The influence of an electric field on a poorly conductive membrane such as
a lipid bilayer is studied theoretically. The unbalanced electric stress created 
by an ionic current across a non-perfectly flat membrane
gives rise to a destabilizing surface energy enhancing undulations. The
deformation of a membrane attached
to a frame and the subsequent force on the frame are derived and the
electrohydrodynamic instability of
a free floating membrane is also studied. We find a most unstable mode of
undulation, of wavelength in
the $\mu m$ range, connected to the crossover between membrane and solvent
dominated dissipations.
\end{abstract}

\pacs{87.16.Dg Membranes, bilayers, and vesicles\\
47.20.-k Hydrodynamic stability\\
87.50.Rr Electric fields\\
submitted to {\em Phys Rev Lett} on June 28 2001}
\vskip1truecm
%

\narrowtext

Due to their low permeability to electrolytes, biological or other lipid
membranes are strongly influenced by applied electric fields. Researchers
are actively investigating phenomena such as
{\em electroporation} (creation of long-lived pores in a lipid membrane
under the action of a
strong electric field)\cite{electroporation,isambert}, in part because of
very promising applications for gene therapy. Similarly, the
{\em electroformation} of
lipid vesicles is a widely used technique to form large unilamellar vesicles
under the action of an electric field\cite{angelova}. Despite its importance,
there is no clear understanding of this experimental tool, which remains
mostly empirical. Many other situations (electrofusion\cite{electrofusion},
electro-injection of macromolecules in vesicles and
cells\cite{incorporation}) require a better understanding of the action of
electric fields on lipid bilayers. This paper addresses this issue
theoretically, and discusses the effect of an electric field on {\em i)}
the force exerted on a rigid frame by an attached membrane, and {\em ii)}
the dynamical instability of a free floating membrane and its relevance for
the
electroformation of neutral liposomes. Before reaching these two topics, we
study the stability of an almost flat membrane under electric field.

The force acting on an interface in an electric field $E$ can be calculated
by evaluating the discontinuity of the Maxwell stress tensor
$\sigma_{ij}=\epsilon(E_iE_j-{1\over 2}E^2\delta_{ij})$ across the
interface\cite{jackson}. We consider an infinite membrane of dielectric constant
$\epsilon_m(\simeq 2\epsilon_0)$ and conductivity $\chi_m^{-1}(\simeq
10^{-6} S/m)$
in a solvent of dielectric constant $\epsilon_1(\simeq 80\epsilon_0)$ and
conductivity
$\chi_1^{-1}(\simeq 10^{-1} S/m)$ (numbers are typical for a lipid membrane
in water, $\epsilon_0$ is the  permitivity of vacuum). A fixed electric field
$E$ (in practice, a fixed electric current $j$) is applied across the membrane,
and we note $E_m$ the electric field {\it inside} the membrane.

If the membrane is {\it perfectly} flat, the electric stress is symmetrically
balanced on both sides of the membrane (i.e., $\sigma_{el} \simeq \epsilon_m E^2_m - 
\epsilon_m E^2_m =  0$), which experiences a mere compression\cite{helfrich}.
As already pointed out in \cite{isambert}, this perfect cancellation of the
net electric stress is however accidental since any membrane curvature leads
to an unbalanced net stress whose order of magnitude can be expressed
as $\sigma_{el} \simeq \epsilon_m E^2_m d/R$, where $d$ is the thickness of 
the membrane and $R$
its local radius of curvature (a curvature of the membrane leads to a
non-uniform surface charge density -
see Fig.1). In the case of a closed vesicle
\cite{isambert}, $E_m$ is given by the potential drop at the vesicle scale
(i.e., $E_m \sim  E R/d$) as most of the electric field goes around the
finite sized object, whereas the current continuity across the (infinite)
membrane dictates the expression of $E_m$ here, namely
$E_m \sim  E {\chi_m / \chi_1}$.
This resulting stress $\sigma_{el}$ tends to {\it enhance} the (small) local
undulations of the membrane $u(r)=\sum_q u_q e^{i q r}$ (Monge representation
in real and Fourier space). After a short transient time
$\tau_e \simeq (\epsilon_m / d) \chi_1 H\simeq 10^{-6}-10^{-5}$s ($H$ is
the distance between the electrodes), the solution of the Poisson equation
(in the limit $\epsilon_m\chi_m^2\gg\epsilon_1\chi_1^2$) yields the following
expression for a given Fourier mode:
$\sigma_{el}=2\epsilon_m (E \chi_m / \chi_1)^2(e^{qd}-1)/(e^{qd}+1)u({\bf r})$. Hence, integrating the work of
the electric stress (in the large wavelength limit $qd\ll1$) yields a net decrease in energy
\begin{equation}
F_{el}=-{\Gamma_{el}\over 2}\int dS(\nabla u)^2\qquad\Gamma_{el}=\epsilon_m\chichi^2 E^2 d
\label{gammae}
\end{equation}
This amounts to  an effective {\it negative} surface tension $\Gamma_{el}$ acting 
on the membrane. In the presence of a lipid reservoir (or at the
interface between two immiscible fluids of different conductivities), this term 
leads to an electric
field induced decrease of the interfacial tension, known as the {\em
electrocapillary\ effect}\cite{capill}. 
On the contrary, for a fixed number of surfactants,
enhancing membrane undulations under electric field builds a concomitant 
mechanical tension because of deviation from the nominal area per molecule.
Hence we expect a tense yet floppy-looking membrane in this case.
For typical values of the electric field used in the
electroporation and electroformation experiments $E\sim10^3 V/m$, the
electrostatic surface tension reaches $\Gamma_{el}=10^{-3}J/m^2$, which is of order the mechanical tension 
needed to rupture a lipid
membrane\cite{electroporation}.

Two consequences of this destabilizing effect are
studied below: {\em i}) a membrane on a fixed frame is deformed
until the electric stress is balanced by an opposing mechanical stress and
{\em ii})
a free membrane undergoes strong undulations under electric field.

\vskip0.3truecm
$\diamond$ \underline{Case of a bilayer attached to a fixed frame}
\vskip0.2truecm
The elastic behavior of a lipid membrane is generally characterized by a
bending modulus $\kappa$ ($\sim 5\times10^{-20} J$), and a stretching
modulus $K_s$ ($\sim 0.1\ J/m^2$)\cite{helfrich2}. Typically, bending a
lipid membrane involves energies of order $1-10 k_BT$,
while stretching it requires much larger energies. To simplify
the description below, we will not include the bending rigidity in the
treatment of the static deformation of the membrane, nor will we include
the thermal fluctuations of the membrane, restricting ourselves to large
electric fields.

We write a linear theory for small
deformations of the membrane $q u_q\ll1$. The membrane is characterized by
its total area $S$, its projected area $S_p$, and its optimum area for
which it is not stretched $S_0$. We use the two small parameters
$\delta\equiv (S_p-S_0)/S_0$ and $\ci\equiv (S-S_p)/S_0\simeq(1+\delta){1\over 2}\sum q^2
|u_q|^2\simeq{1\over 2}\sum q^2
|u_q|^2$. The electric-field-induced undulation term involves the parameter 
$\ci$ only
(Eq.(\ref{gammae})), and the stretching energy involves the total increase
of area with respect to $S_0$:
\be
F=S_0\left({1\over 2}K_s(\delta+\ci)^2-\Gamma_{el}\ci\right)
\label{en1}
\ee
Note that the electric-field-induced undulation term ($\Gamma_{el}$) is 
quadratic in the small
quantity $q u_q$, while the stretching term ($K_s$) corresponds to an
expansion to the fourth order, consistent with the large value of the ratio
of the stretching over the electrostatic parameters $\beta\equiv
K_s/\Gamma_{el}>100$.

The minimization of this energy with respect to the out-of-plane membrane
deformation $\ci$ leads to an equilibrium class of membrane shapes
corresponding to a total area difference $\Delta_S$
\be
\Delta_S=\delta+\ci_{eq}={\Gamma_{el}\over K_s}\sim 10^{-2}
\label{eq1}
\ee
The energy of the membrane only is $F_m={1\over 2} S_0{\Gamma_{el}^2\over K_s}$,
and the energy of the system, including the electrostatic energy, is
$F=F_m+F_{el}=-{1\over 2} S_0{\Gamma_{el}^2\over K_s}+S_0\Gamma_{el}\delta$.

The surface tension of the lipid membrane at equilibrium is given by
$\gamma=\partial F_m/\partial S$ with
$S=S_0(1+\delta+\ci)$ 

At equilibrium, the mechanical tension is equal, but
opposite in sign, to the
electrostatic
energy per unit area: $\gamma=\Gamma_{el}$. The building of a mechanical
tension in a membrane under electric field is due to an increase of its
area. 

Beyond the electroporation threshold, holes in the membrane reduce
the mechanical tension by decreasing the membrane area and strongly
increasing the membrane electrical conductivity $1/\chi_m$. The effect of
the electric field could be measured via the setup depicted in Fig.2. The
force measured by the spring is $f=\partial F_{tot}/\partial L_p=\Gamma_{el}
L_y$ ($\sim 10^{-3}N/m$), where $L_p$ and $L_y$ are the membrane projected
size in the direction of and perpendicular to the spring respectively. The
membrane pulls on the spring ($f>0$)
if $\ci>-\delta$. The force in the
absence of field is simply: $f_0=K_s L_y\delta$.

It is clear from Eq.(\ref{eq1}) that the balance between electric-field-induced
undulation and membrane stretching
does not select a particular equilibrium membrane shape, as 
both effects depend on
the global increase of membrane area only. Including the bending
energy of the membrane selects the shape of lowest curvature, namely
the first harmonic $q_1=\pi/L_p$. The energy gap between the different
harmonics is however small (of order $k_BT$) for $q \lambda_\kappa<1$, where
$\lambda_\kappa=2\pi\sqrt{\kappa/\Gamma_{el}}$. For large electric fields
($E\sim10^3 V/m$), $\lambda_\kappa$ is very small ($\sim 10 nm$). For small
electric fields, the thermal fluctuations dominate both the membrane shape
and tension. We postpone the study of the interesting crossover between
these two limits to a future publication.

There are some dynamical issues connected to force measurements on lipid
membranes (Fig.2). Lipid membranes spread on a frame are generally connected
to the frame by meniscii which are large compared to the membrane thickness,
and from which lipid molecules may flow toward the deformed membrane (a
phenomenon reminiscent of the Marangoni
 effect\cite{marangoni}). In this case, the membrane elastic stress consequent to the action of an electric field is rapidly released by migration of lipid molecules, and might not be detectable experimentally. The kind of force measurement depicted  on Fig.1 should however be feasible on polymeric membranes, which have shown sensitivity to electric field as well\cite{polymeric}, and for which the
migration of molecules along the membrane, if any, is very slow. 
An alternative experimental setup for lipid membranes involves measurements on lipid membrane strongly bound to a solid substrate, which are expected to exhibit a fairly slow lateral dynamics.


\vskip 0.3truecm
$\diamond$ \underline{Instability of free floating membranes under E-field}
\vskip0.2truecm
We now study the case of a free floating membrane under an electric field.
This study is partly motivated by the experimental technique of
electroformation of vesicles, which allows a controlled swelling of an
electrode deposit of lipids to form vesicles of fairly well controlled
sizes under electric field\cite{angelova}. Interestingly, this technique
works for charged and nonionic lipids, although the optimal conditions
vary widely with the nature of the lipids.

The dynamics of a free membrane under electric field can be decomposed into
two mechanisms. {\it i}) The normal deformation of the membrane (described
in the previous section), which creates a tension $\gamma$ in the membrane and saturates
when $\gamma = \Gamma_{el}$.
{\it ii}) The lateral sliding (contraction) of the membrane, in an attempt to release
this tension.
The characteristic time for the normal deformation for a given undulation
mode of wavelength $\lambda$ is $\tau_\perp\sim\eta\lambda/\Gamma_{el}$ ($\sim
10^{-6}s$ for $\lambda=1\mu m$), while the sliding motion involves the
whole membrane $\tau_\parallel\sim \eta L/\Gamma_{el}$
($\sim 10^{-3}s$ for $L\sim 1mm$). The two mechanisms occur at very different timescales,
and can be treated separately.

A thorough treatment of the undulation modes of a film immersed in a solvent
can be found in the
literature\cite{undul1}. The electrohydrodynamic
instability of a layer of non-conducting fluid between two semi-infinite
conducting fluids has been studied in \cite{michael}, where special
attention is given to the peristaltic deformation modes (the two interfaces
undulating in antiphase), as these modes lead to the destruction of
the film when the two interfaces make contact. In the case of a lipid
membrane, these
peristaltic modes are suppressed because of the very low
compressibility of the film. We study below the bending instability (interfaces undulating in phase) of the membrane.

{\em i}) The normal displacement of the membrane involves viscous
dissipation in and around the membrane \cite{seifert,evans}. There are three
main dissipation mechanisms, namely the dissipation in the solvent
(viscosity $\eta=10^{-3} Pa.s$), which dominates the dynamics of large
wavelengths deformation, the
friction between the two monolayers (friction coefficient $b_{fr}=10^8
Pa.s/m$), dominant at intermediate wavelengths,
and the membrane surface
dissipation (surface viscosity $\mu=10^{-10} Pa.s.m$). For undulating 
membranes, the latter mechanism is relevant at very small wavelength of 
the order of the bilayer thickness $d=5nm$ only, and will
be neglected in what follows.
We present below a simplified description of the interplay between external
and internal dynamics. For a thorough treatment of membrane dynamics, see
Seifert\cite{seifert}.

Neglecting inertia, a normal deformation of the membrane (of typical
lateral size $2\pi/q$ and typical velocity $\dot u_q$) creates a motion in
the surrounding fluid which propagates to a distance $\sim1/q$. The curvature of the membrane leads to a velocity difference of
order $\delta v=q d \dot u_q$ between the two monolayers. The power
dissipated by viscous effect around and in the membrane can be written
respectively:
\begin{eqnarray}
P_\eta=\eta\int dV \left(\nabla v\right)^2=S\eta \sum_q q\dot u_q^2\cr
P_{fr}=b_{fr}\int dS \left(\delta v\right)^2=Sb_{fr} \sum_q d^2q^2\dot u_q^2
\label{diss1}
\end{eqnarray}
This dissipated power must compensate the power stored in the membrane
$P_m=\partial_t F$ (the energy $F$ is given by Eq.(\ref{en1})). This
condition leads to
an evolution equation for each deformation mode:
\begin{eqnarray}
(\eta q+b_{fr} d^2 q^2)\dot u_q(t)=\hskip3truecm\cr
\left[\Gamma_{el} q^2\left(1-\beta(\delta(t)+\ci(t))\right)-\kappa
q^4\right]u_q(t)
\label{bal1}
\end{eqnarray}
with $\beta\equiv K_S/\Gamma_{el}>100$. The left hand side of this equation
describes the viscous dissipation in and around the membrane and the right
hand side consists of the electric-field-induced undulation term ($\Gamma_{el}$) 
with a stretching saturation ($\beta$ term). Note that the bending
rigidity of the membrane has been added to the membrane energy ($\kappa$
term), for it is mandatory for the description of the small
wavelength deformations $q>\sqrt{\Gamma_{el}/\kappa}$. This equation is
non-linear, since the saturation involves the total increase of area
$\ci={1\over 2}\sum_q q^2 |u_q|^2$. Thanks to the different time scales for
normal and
lateral motions, the projected area of the membrane $S_p(t) \propto \delta(t)$ can be
considered as constant for the short time evolution.

This non-linear equation can be solved, but for our purpose,
it is sufficient to consider its linearized form, and to treat the
saturation dynamics separately. The short time evolution is described by the linear
equation
\be
(\eta q+ b_{fr} d^2 q^2) \dot u_q=(\Gamma_{el} q^2 -\kappa q^4) u_q
\label{dispsimple1}
\ee
The amplitude of a given Fourier mode has a time evolution $u_q(t)\sim
e^{\alpha_q t}$ with
\be
\alpha_q={\Gamma_{el}\over\eta}q\left({1-(q/qÑ{stat})^2\over1+q/q_{dyn}}\right)
\label{rate}
\ee
with the two characteristic wavevectors:
\be
q_{stat}=\sqrt{\Gamma_{el}\over\kappa}\sim 5.10^7 m^{-1} \quad q_{dyn}={\eta \over
b d^2} \sim5.10^5 m^{-1}
\label{qcqr}
\ee
The evolution rate presents a sharp maximum at
\be
q^*=\left(q_{dyn} q_{stat}^2\right)^{1/3}= 8.10^6\ m^{-1}
\label{qstar}
\ee
for $q_{stat}\gg q_{dyn}$. This defines a particular lengthscale which grows exponentially faster than
the others: $\lambda^*=2\pi/q^*\sim1\ \mu m$. The corresponding growth rate is 
$\alpha_q^*\sim{\Gamma\over b d^2}\sim 5.10^5 s^{-1}$. The
evolution saturates similarly for all lengthscales when
$\beta\left[\delta(t)+\ci(t)\right]=1$, at which point a
mechanical tension $\Gamma_{el}$ is established, leading to a contraction of the membrane.

{\em ii}) The lateral motion of the membrane occurs at an almost constant
surface tension, because the time needed to build up the tension
$\tau_\perp$ is much smaller than the time $\tau_\parallel$ over which it
could be released. Since the contraction of the membrane involves solvent flow over large lengthscales ($mm$), inertial effects must be included\cite{landau}. The flow
created by the lateral motion of the membrane (of size $L$) propagates over
a size $L_z= L/(\sqrt{1+L^2/(\nu t)})$ ($\nu=\eta/\rho$ is the kinematic
viscosity $\sim 10^{-6} m^2/s$ for water - $\rho$ is the density of water).
The comparison of the power
dissipated by the sliding motion of velocity $\dot L$: $P_{diss}=\eta\int
dV(\nabla v)^2=\eta L^2 \dot L^2/L_z$, with the power stored in the
membrane $P_{stor}=\Gamma_{el} L\dot L$, leads to the dynamical equation for the
membrane size:
\be
\dot L=-{\Gamma_{el}\over \eta}{L_z\over
L}=-{\Gamma_{el}\over\eta}{1\over\sqrt{1+L^2/(\nu t)}}
\label{dynslide}
\ee

The short time evolution ($t<L_{t=0}^2/\nu\sim 1s$) is dominated by the
diffusion of the solvent flow. Assuming that $L_{t=0}=L_0$ (No stretching
without electric field), the evolution equation for short time is
\begin{eqnarray}
L(t)=L_0\sqrt{1-4/3\sqrt{t^3/(\tau_\Gamma^2\tau_\nu)}}\qquad\cr
\tau_\Gamma\equiv\eta L_0/\Gamma_{el}\simeq 10^{-3}s\qquad
\tau_\nu=L_0^2/\nu\simeq 1s
\label{Ltslide}
\end{eqnarray}
from which emerges a characteristic time scale
$\tau_{slide}=\left(\tau_\Gamma^2\tau_\nu\right)^{1/3}\sim 10^{-3}s$. This
result
is reminiscent of the dynamics of bursting of a soap film in a viscous
environment\cite{jf}. The evolution of the variable $\delta=L_p/L_0-1$
which appears in Eq.(\ref{bal1}), is given by
$\delta(t)\simeq2/3(t/\tau_{slide})^{3/2}$.


In many practical situations, fixed
boundaries or walls (the electrodes in the case of the electroformation of
vesicles) may
strongly modified the hydrodynamics. The solvent viscous dissipation (Eq.(\ref{diss1})) is modified by the presence of a wall at
a distance $h$ from the membrane, and becomes $P_{\eta,h}=S\eta\sum_q\dot
u_q^2/(q^2h^3)$ for $qh<1$. The left hand side of Eq.(\ref{dispsimple1}) is
modified accordingly, and shows an optimal wavevector $q^*_h\sim
\sqrt{q^*/h}$ where $q^*$ is given by Eq.(\ref{qstar}). For $h=10 nm$, the
fastest growing wavelength is of order $2\pi/q^*_h\sim0.1\mu m$. The sliding
 motion is strongly affected by the wall. For times larger than $h^2/\nu$,
the time dependent length $L_z$ should be replaced by the distance $h$ in
Eq.(\ref{dynslide}), and the dynamics is described by
$L(t)=L_0\sqrt{1-t/\tau_h}$ with $\tau_h=\tau_\Gamma L_0/h\sim 10^2 s$
instead of $\tau_{slide}=10^{-3}s$ (note that solvent permeation through lipid membrane becomes relevant near a wall)!

To conclude this paper, we would like to propose that the
electrohydrodynamic instability described above may play an important role in
the first stage of the electroformation of liposomes. A remarkable feature of this technique is that it produces vesicles of fairly well defined size. We find a fastest growing undulation mode of wavelength
in the $\mu m$ range (Eq.(\ref{qcqr},\ref{qstar})). This mode might be the precurssor of large scale deformations of the membrane which, after a complex process partly sketched in Ref\cite{angelova} and involving coalescence of neighboring blisters, lead to the
formation of closed vesicles (the size of which can reach $50\ \mu m$ for nonionic
lipids\cite{dimeg}). Future developments will include the
treatment of small pores which are expected to be present in a membrane
under tension, and their influence on both the membrane electrical
conductivity (hence $\Gamma_{el}$) and dynamics (solvent permeation through the
membrane).

We would like to thank A. Johner, M. Angelova, R. Bruinsma, J.F. Joanny, A. Ajdari and J.B Fournier for stimulating discussion and useful comments.

\newpage

\vskip0.5truecm
\centerline{\epsfxsize=12truecm \epsfbox{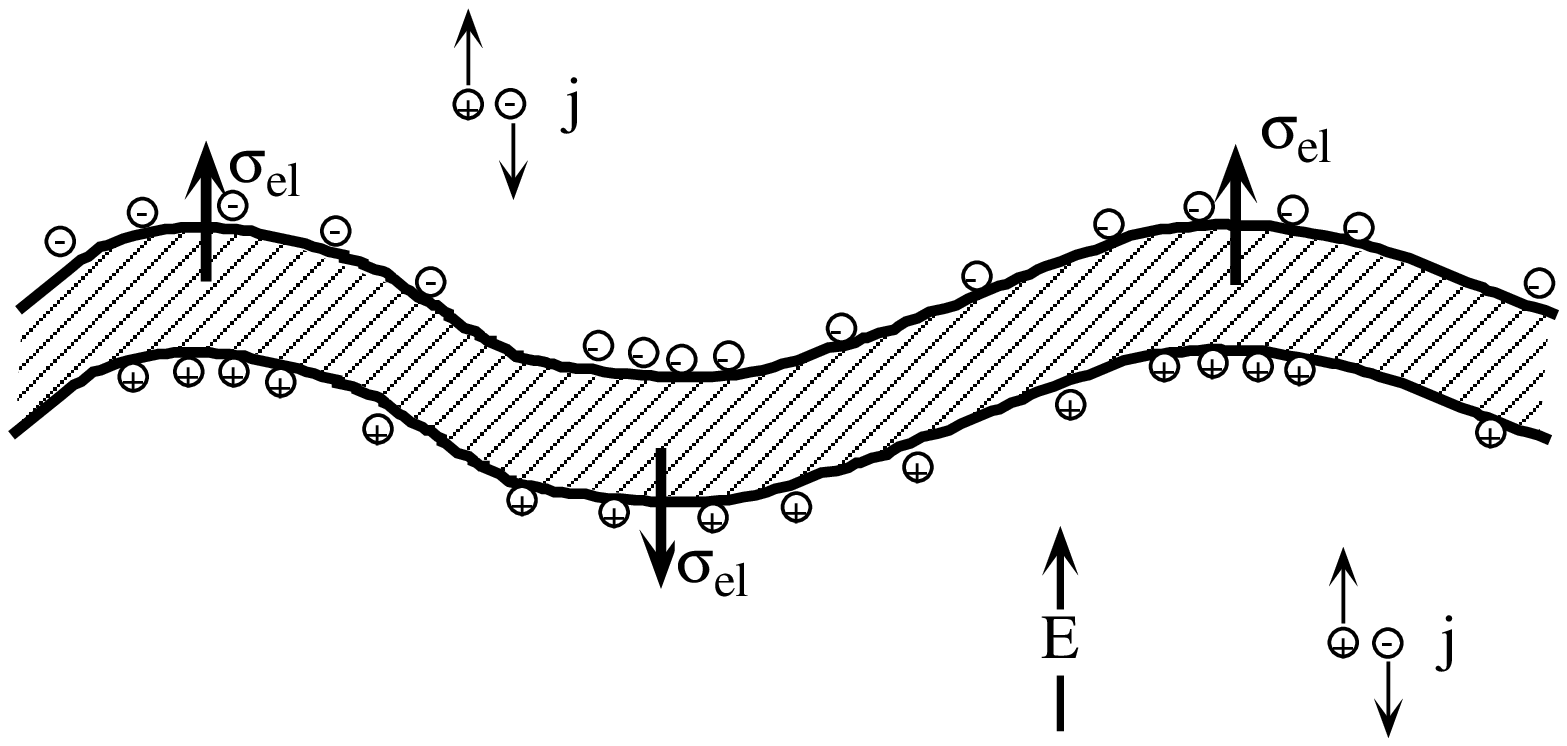}}
\vskip2truecm\centerline{Fig.1 Net accumulation of conduction charges 
near}
\centerline{\footnotesize  a curved lipid membrane under electric field}
\vskip0.5truecm

\vskip6truecm

\vskip0.2truecm
\centerline{\epsfxsize=12truecm \epsfbox{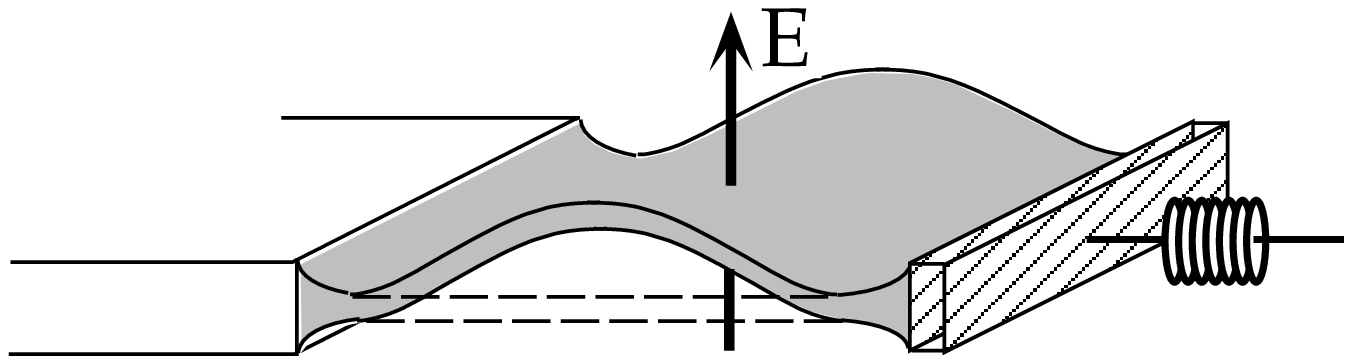}}
\vskip2truecm\centerline{Fig.2. Possible experimental setup for the force
measurement}
\vskip0.5truecm

\end{document}